\documentclass[twocolumn,amssymb,nobibnotes,aps,prl,superscriptaddress]{revtex4-2}
\usepackage{graphicx}
\usepackage{textcomp}
\usepackage{amsmath}
\usepackage{commath}
\usepackage{color}
\usepackage[dvipsnames]{xcolor}
\usepackage{textcomp}
\usepackage{amssymb}
\usepackage{matlab-prettifier}
\usepackage{siunitx}

\newcommand{\recAng}{\theta_r}
\newcommand{\conAng}{\theta_c}

\newcommand{\resDom}{\Omega_R}
\newcommand{\prisDom}{\Omega_P}

\newcommand{\freq}{f}

\newcommand{\ts}{\textsuperscript}
\newcommand{\lego}{\ts\textregistered{}}
\newcommand{\profile}{\delta}

\begin{document}
	
	\title{Programming droplet motion using metamaterials}
	\author{Mohammad Charara}
	\affiliation{Department of Civil, Environmental, and Geo- Engineering, University of Minnesota, Minneapolis, Minnesota 55455, USA}
	\author{Zak Kujala}
	\affiliation{Department of Mechanical Engineering, University of Minnesota, Minneapolis, Minnesota 55455, USA}
	\author{Sungyon Lee}
	\email{sungyon@umn.edu}
	\affiliation{Department of Mechanical Engineering, University of Minnesota, Minneapolis, Minnesota 55455, USA}
	\author{Stefano Gonella}
	\email{sgonella@umn.edu}
	\affiliation{Department of Civil, Environmental, and Geo- Engineering, University of Minnesota, Minneapolis, Minnesota 55455, USA}
	
	\begin{abstract}
		Motion control of droplets has generated much attention for its applications to microfluidics, where precise 
		control of small fluid volumes is an imperative requirement. 
		Mechanical vibrations have been shown to be effective at inducing controllable depinning, and activation of different drop motion regimes. However, existing vibration-based strategies involve establishing homogeneous rigid-body dynamics on the substrate, and therefore lack any form of spatial heterogeneity and tuning. Addressing this limitation, metamaterials provide an ideal platform to achieve spectrally and spatially selective drop motion control, which leverages their ability to attenuate vibrations in selected frequency bands and in selected regions of a substrate. In this work, we illustrate the potential of metamaterials-based drop control by experimentally demonstrating a variety of drop motion capabilities on the surface of metaplates endowed with locally resonant stubs. The experiments leverage the design versatility of a LEGO\lego{} component-enabled reconfigurable design platform and laser vibrometry measurements with high spatial resolution.
	\end{abstract}
	
	\maketitle
	
	Research aimed at controlling the motion of fluid droplets has blossomed in recent years, inspired by progress in microfluidics, where precise control of small liquid volumes enables a variety of chemical and biological processes~\cite{stone2004engineering, mugele2005electrowetting, squires2005microfluidics, seemann2011droplet, bintein2019grooves}. Triggering 
	drop motion on a 
	substrate requires breaking the symmetry of the drop. This happens when the \emph{contact angle} $\conAng$ between drop and substrate increases past the \emph{advancing angle} $\theta_a$ on one side of the drop and decreases below the \emph{receding angle} $\recAng$ on the other. $\theta_a$ and $\recAng$ are intrinsic properties of the fluid-drop system, 
	and their difference, called \emph{contact angle hysteresis} (CAH)~\cite{gao2006contact, eral2013contact}, can be interpreted as an energy barrier \emph{pinning} a drop to a surface. Strategies to overcome CAH and control drop motion include chemical~\cite{ huang2014numerical, varagnolo2014tuning, lin2018tuning, lian2020directional} and mechanical~\cite{ buguin2002ratchet, dupeux2014propulsion, sheng2011directional, lian2020directional} doctoring of substrates, surface acoustic waves~\cite{yeo2009ultrafast, baudoin2012low, bussonniere2016dynamics, noori2021surface, brunet2022unstationary}, and vibrations~\cite{lian2020directional,daniel2002rectified, daniel2005vibration, dong2006lateral, celestini2006vibration, brunet2007vibration, brunet2009motion, noblin2009ratchetlike, john2010self, sartori2015drop, ding2018ratchet, sartori2019motion, costalonga2020directional, deegan2020climbing}
	
	Vibration-induced drop motion is typically achieved by connecting a substrate to a shaker or speaker, which imparts to the \textit{entire} surface a harmonic \textit{rigid-body} motion at a frequency $\freq$. Under certain frequency and amplitude conditions, the contact line of a deposited drop can \emph{depin} from the substrate, leading to drop motion. In the case of a flat surface, symmetry can be broken via an inclined actuation source~\cite{daniel2005vibration, dong2006lateral, costalonga2020directional}. 
	On an inclined surface, however, gravity is sufficient to provide the necessary asymmetry. Here, drops have been shown to exhibit a variety of vibration-controlled motion regimes, including stop-and-go downward sliding, stationary depinning, and even upward climbing, depending on the vibration characteristics and inclination angle \cite{brunet2007vibration, brunet2009motion, sartori2015drop, ding2018ratchet, deegan2020climbing}. Standard rigid-body vibration strategies generate a \textit{homogeneous} landscape of vibrations, 
	whereby a drop experiences nearly the same acceleration forces across the substrate. 
	For applications that require \textit{spatially heterogeneous} control, it is imperative to devise methods that allow us to activate (i.e., excite) drops differently depending on their location. 
	Existing methods to spatially diversify the drop response often involve doctoring the substrate, 
	thus introducing an element of irreversibility. 
	The challenge is to devise a seamless strategy to achieve 
	\textit{spatial selectivity} 
	without requiring any modification of the 
	surface, while displaying a high degree of reconfigurability.
	
	\begin{figure*}[t]
		\centering
		\includegraphics[width=176truemm]{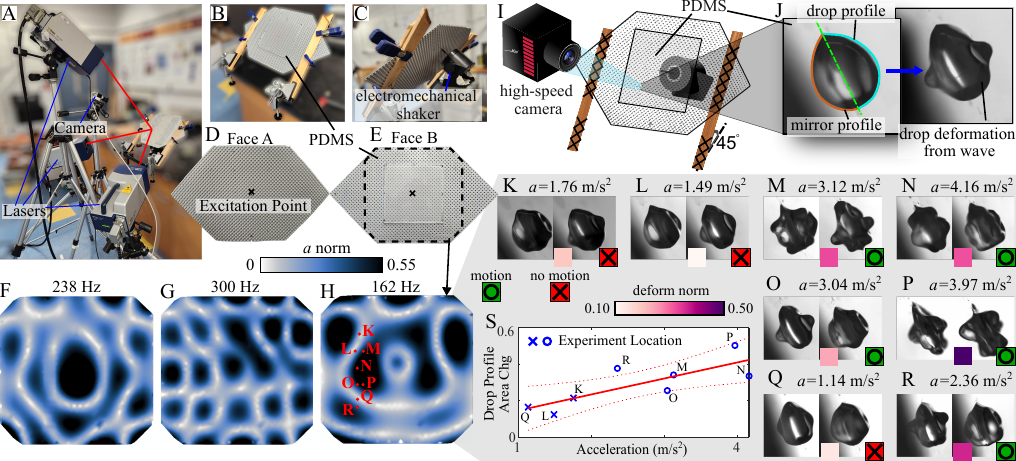}
		\caption{(A-C) Experimental setup for testing the \emph{pristine} LEGO\lego{} baseplate, showing the 3D laser Doppler vibrometer and shaker. (D) face A and (E) face B of the baseplate, with a PDMS patch on face B. Sample normalized acceleration fields at (F) $238 \, \textrm{Hz}$, (G) $300 \, \textrm{Hz}$, and (H) $162 \, \textrm{Hz}$, reconstructed via laser Doppler vibrometry under harmonic excitation, highlighting patterns of peaks, valleys, and nodal lines. 
			(I) Schematic of setup for high-speed camera video collection of vibrating drops. 
			(J) Frame captures from a sample video, showing a drop on the PDMS surface and its deformed configuration under vibration. (K-R) Profiles of drops experiencing excitation at $162 \, \textrm{Hz}$, with 
			corresponding locations marked on 
			field H; the white-to-purple color scale quantifies the deformation, and the green-red binary label denotes motion vs. no motion of the drop. (S) Positive correlation between local plate acceleration and drop deformation.} 
		\label{fig:pristPlate}
	\end{figure*}
	
	A possible solution comes from the realm of \emph{elastic metamaterials}, architected networks of structural elements whose geometry, spatial arrangement, and connectivity yields emergent ensemble properties that transcend those of the individual components \cite{smith2000negative, zhang2011broadband, babaee20133d, rafsanjani2016bistable, meeussen2020topological, zunker2021soft}. 
	They enjoy special wave manipulation capabilities, including notably the ability to generate \emph{bandgaps} (BG) --- frequency intervals in which waves are evanescent --- thus acting 
	as mechanical filters that attenuate vibrations in desired frequency ranges \cite{ruzzene2005directional, krushynska2017coupling, nanda2018tunable, elmadih2021metamaterials}. A subset of these, \emph{locally resonant metamaterials} (LRMs), specifically, rely on 
	resonant mechanisms internal to their unit cells to open low-frequency, subwavelength bandgaps~\cite{liu2000locally, lemoult2013wave, liu2018broadband, celli2019bandgap, muhammad2020bandgap}. 
	A popular 
	class of LRMs is represented by stubbed plates, i.e., plates featuring stub- or pillar-like resonators~\cite{wu2008evidence, oudich2010sonic, oudich2011experimental, yao2011low, celli2015manipulating}.  Here, a convenient feature is that the stubs can be incorporated 
	on one face of the plate, leaving the opposite 
	face intact, 
	thus not affecting its fluid-substrate interaction properties. 
	
	When an LRM is embedded in a conventional elastic medium, the region featuring resonators can be isolated from incoming vibrations 
	with frequencies 
	falling in the bandgap. This results in the ability to program spatially selective patterns of vibration 
	by simply engineering the spatial layout of the resonators. Harnessing these capabilities, in this work we present a metamaterials-enabled strategy to achieve \emph{spatially-selective} drop motion control. This approach leverages the interplay between three conceptual pillars: 
	1) drops are sensitive to changes in substrate vibration amplitude, switching between rest and motion in response; 
	2) resonant mechanisms can open bandgaps that attenuate vibrations in the frequency intervals of interest for drop motion; 3) it is possible establish heterogeneous patterns of vibration isolation via spatially non-uniform resonators layouts. An agile framework to assemble LRM prototypes using LEGO$^{\circledR}$ baseplates and bricks was first introduced by Celli and Gonella~\cite{celli2015manipulating}, and later used to explore 
	bandgap broadening~\cite{celli2019bandgap} and 
	topological pumping~\cite{rosa2021exploring}. Here, the baseplate acts as a wave-carrying medium and the bricks work as pillar-like resonators. The ease of resonator reconfigurability allows for on-the-fly tuning of the LRM functional frequency, and rearranging the bricks allows for on-demand variation of the spatial selectivity. 
	In the context of surface-drop interaction, this translates into the possibility of allowing or disallowing drop motion in different regions. The remainder of the article provides an experimental demonstration of this idea 
	and a first glimpse at the vast landscape of unconventional drop control 
	functionalities that it enables.
	
	The problem of vibration-based drop motion on a \textit{flexible} plate immediately introduces an additional layer of complexity compared to the \textit{rigid}. 
	When subjected to a 
	point excitation, a flexible plate develops a spatially non-uniform response, 
	characterized by peaks and valleys separated by 
	nodal lines. The spatial periods depend on the frequency, shape of the domain, and boundary conditions. In the context of drop motion, this built-in heterogeneity in the plate response results in 
	an intrinsic spatial variability of the substrate conditions experienced by droplets. Therefore, before considering a plate with resonators --- where the resonators 
	introduce an additional spatial selectivity that eventually overrides the intrinsic heterogeneity --- it is important to characterize the response of a \emph{pristine} plate, and explore how drops 
	respond to it. 
	
	We take as our reference pristine plate a simple LEGO\lego{} board [acrylonitrile butadiene styrene (ABS)], and characterize its response to point excitation. The testing setup, detailed in Fig.~\ref{fig:pristPlate}A-C, shows the plate inclined at $45^\circ$, with face A displaying the array of built-in studs for brick connection (Fig.~\ref{fig:pristPlate}D), and face B featuring their negatives (Fig.~\ref{fig:pristPlate}E), which result in an undesired pattern of surface dips. To establish 
	flatness on face B, we coat a rectangular region with a thin layer of polydimethylsiloxane (PDMS) (Silgard 182 -- see SM), for later use 
	as a substrate in 
	experiments. The PDMS layer also endows 
	the surface with higher hydrophobicity to promote the onset of drop motion. An electromechanical shaker applies an out-of-plane excitation through a stinger at the plate mid-point, ensuring 
	overall symmetry in the response (Fig.~\ref{fig:pristPlate}C). 
	We experimentally reconstruct the out-of-plane response field through 3D laser Doppler vibrometry scans of the vibrating plate surface (Fig.~\ref{fig:pristPlate}A -- see SM). We prescribe a broadband excitation of 1--400 \unit{\hertz} and collect velocity data (later converted to acceleration) at the scan points 
	of a dense 1294-point 
	scanning grid on the surface of the plate. Finally, discrete Fourier transform (DFT) 
	yields spectral amplitude maps (i.e., wavefields): 
	Figs.~\ref{fig:pristPlate}F--H show results at sample frequencies $238, \, 300, \, $ and $ 162 \, \textrm{Hz}$, respectively, with darker (brighter) colors denoting higher (lower) acceleration. 
	
	A drop deposited on a plate 
	is expected to feel a strength of excitation depending on its location, 
	yielding commensurate levels of activation, ranging from light to heavy deformation, and even drop breakup. 
	Inclining the plate generates gravity-based asymmetry in the drop profile, and a deformation of the drop sufficiently pronounced to meet the contact angle criterion can result in sliding down the incline. 
	To capture the transfer of energy from the plate to the drops and document the conditions under which this translates into drop motion, we probe the drop-plate interaction in different regions. We deposit 20 \unit{\micro \liter} drops (water) at eight locations 
	(labeled K--R in Fig. \ref{fig:pristPlate}H), and excite the plate harmonically at $162 \, \textrm{Hz}$, 
	a frequency 
	that yields a heterogeneous acceleration field 
	in the K-R observation window; droplet locations are chosen to span a range of accelerations. 
	The drop profile evolution over an excitation cycle is captured using a high-speed camera (Photron Mini AX200 -- see SM) at a 2000 fps sampling rate. Three droplets are tested at each location 
	to ensure consistency. For each location, we extract the deformation at the two extrema of the cycle, as shown in Fig. \ref{fig:pristPlate}K--R. From image segmentation of the profile, 
	we quantify the drop deformation 
	at the peak of the cycle, and use this as a measure of drop \emph{activation}, with intensity marked by a 
	white-purple color bar. Moreover, we classify each location binarily, marking it \emph{red (X)} or \emph{green (O)} according to whether the drop stayed pinned or moves, respectively, during the excitation.

	The 
	images reveal that the drop deformation 
	correlates with the acceleration experienced by the surface. This observation is quantitatively supported by the graph in Fig. \ref{fig:pristPlate}S, which plots the relative change in drop profile area against the plate acceleration at the drop location, where the linear fit asserts a positive correlation between local plate acceleration and drop deformation. 
	Additionally, a drop deposited at locations K, L and Q (M--P and R), which are regions of low (high) acceleration, experiences no motion (motion), suggesting an overall positive correlation between drop activation and onset of motion.
	However, we report intermediate situations (e.g., location O in our test), where motion is triggered under limited deformation levels. 
	An important lesson 
	about the complexity of working with flexible substrates emerges from these results. The substrate inherently brings about a spatial heterogeneity intrinsic to its modal response, which is transferred to the drops in the form of a heterogeneous landscape of activation. In trying to program drop motion, we cannot ignore this built-in heterogeneity, but only contrast it with a spatial selectivity imposed \textit{by design} via our control strategy. 
	
	\begin{figure*}[t]
		\centering
		\includegraphics{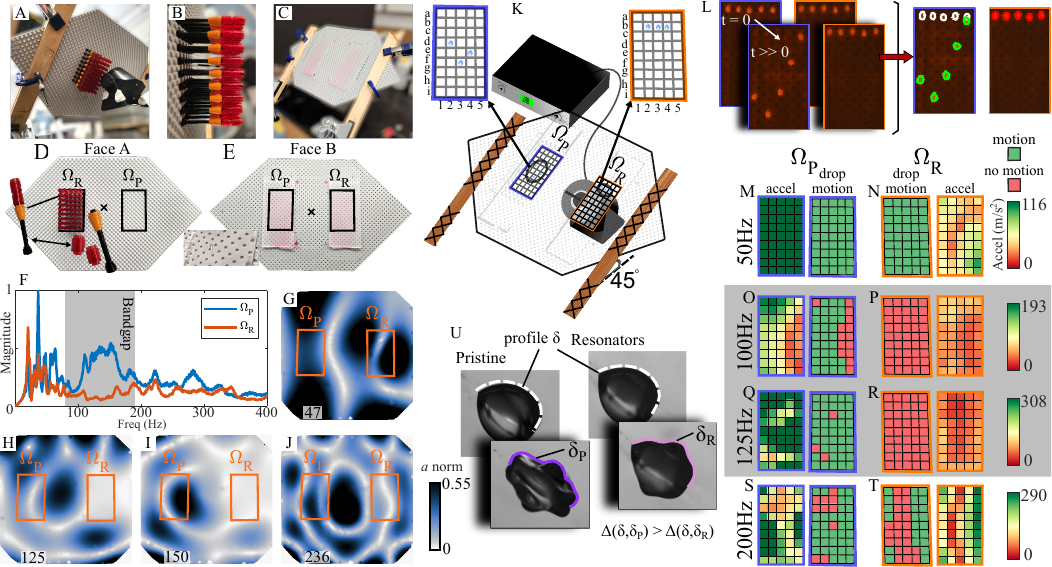}
		\caption{(A) Rear, (B) side, and (C) front views of  LRM metamaterial plate setup, with side view showing details of the resonator array. (D) Face A and (E) face B of the plate, showing a 9x5 array of resonators arranged inside $\resDom$ and a pristine $\prisDom$ on face A, and two PDMS patches placed on face B for drop motion experiments. (F) Frequency response function of the plate from laser Doppler vibrometry measurements, obtained by averaging responses at points inside $\prisDom$ (blue) and $\resDom$ (orange), showing a bandgap in the 90-190 \unit{\hertz} range. (G--J) Acceleration wavefields for frequencies (G) below the BG ($47 \, \textrm{Hz}$), (H,I) inside the BG ($100 \, \textrm{Hz}$ and $125 \, \textrm{Hz}$), and (J) above the BG ($236 \, \textrm{Hz}$), showing attenuation inside $\resDom$ for frequencies in the BG. (K) Setup for drop motion experiments showing $\prisDom$ (blue outline) and $\resDom$ (orange outline) divided into 9x5 subdomains. (L) Drop motion captures showing stationary drops before vibration ($t=0$), and moving for $t>>0$. Drop motion is classified binarily: drops that have slid by the end of the vibration time are labeled in green and drops that remain pinned in red. (M--T) Outcome of drop motion assessment at different frequencies, with green-red pixels in the subdomain grids to distinguish regions where drops initially deposited move from others where they stay pinned: (M,N) results at $50 \, \textrm{Hz}$ [below the BG]; (O,P) results at $100 \, \textrm{Hz}$ and (Q,R) $125 \, \textrm{Hz}$ [both inside the BG];  (S,T) results at $200 \, \textrm{Hz}$ [above the BG]. Each grid is accompanied by experimentally-measured plate acceleration maps, specifying the maximum acceleration at each subdomain, with high (low) acceleration in green (red). (U) High speed camera captures of a drop deforming in response to a vibration at $125 \, \textrm{Hz}$ (inside the BG), with and without resonators. The undeformed drop profile $\profile$ is distorted to the deformed $\profile_P$ and $\profile_R$ in the $\prisDom$ and $\resDom$, respectively.}
		\label{fig:metaPlate}
	\end{figure*}
	
	Having characterized the pristine plate, we shift our attention to an LRM plate 
	with pillar-like resonators. 
	Our LEGO\lego{} platform, pictured in Fig.~\ref{fig:metaPlate}A-C, allows for 
	a modular and reconfigurable LRM framework. On the 
	baseplate, we identify two rectangular regions, symmetrically located about the vertical axis: $\prisDom$, which we leave pristine, and  $\resDom$, which features on face A 
	a $9 \times 5$ array of rod-brick resonators with conical and cylindrical tip masses (Fig.~\ref{fig:metaPlate}D). The 
	position 
	of the cones along the rods, and the quantity of cylinders can be varied to tune the resonant frequency of the resonators and, consequently, the onset of the opened bandgap~\cite{celli2019bandgap}. 
	The corresponding $\prisDom$ and $\resDom$ regions of face B (Fig. \ref{fig:metaPlate}E) are coated with a thin layer of PDMS that provides a smooth 
	substrate for drop motion. 
	A signature feature of an LRM-enabled strategy is that we can \textit{arbitrarily} select the regions where we want to control the plate response and its effect on drop motion. 
	Specifically, LRMs with \textit{localized} patches of resonators 
	attenuate vibrations 
	only within regions that contain resonators, as long as the driving frequency is within the BG. 
	Accordingly, a drop placed on 
	$\resDom$ is expected to experience attenuated vibration, for $\freq$ inside the BG, 
	thus undergoing small deformation and staying pinned. In contrast, a drop deposited on $\prisDom$ is expected to experience the amplitude of response of the pristine plate (involving large acceleration, unless falling at a nodal line) 
	thus experiencing significant deformation and triggering motion, even when we drive the plate within the BG. 
	
	In designing and tuning the resonators, we balance two competing requirements. The first is the need to approach as closely as possible the frequency of the fundamental pumping and rocking mode of a 20 \unit{\micro \liter} drop (water)~\cite{celestini2006vibration, sartori2015drop} 
	to promote drop activation~\cite{brunet2007vibration, brunet2009motion}. This requirement translates to 
	the need to design resonators with low resonant frequencies, which often means long and slender rods with heavy tip masses~\cite{celli2015manipulating}. The second is the need to maintain reasonable design spec bounds, 
	whereby the resonators cannot be impractically long  or heavy. 
	The pillars shown in Fig. \ref{fig:metaPlate}A--C are the outcome of this trade-off exercise. 
	
	We characterize the bandgap ascribable to the resonators via laser Doppler vibrometry measurements. 
	We excite the LRM 
	with a broadband signal (1--400 \unit{\hertz}), and collect a frequency response curve at every scan point on the surface (see SM).
	Fig.~\ref{fig:metaPlate}F shows the averaged frequency response inside $\prisDom$ (blue) and $\resDom$ (orange). The BG, 
	identified as the gap between the curves, spans roughly the 90--190 \unit{\hertz} range. Interpolating the frequency spectra acquired at points of the dense scanning grid defined over the entire surface, we can also reconstruct spectral acceleration maps at selected frequencies to document differences in response patterns between $\prisDom$ and $\resDom$. 
	Acceleration fields are shown for frequencies below ($47 \, \textrm{Hz}$, Fig.~\ref{fig:metaPlate}G), inside ($125 \, \textrm{Hz}$ and $150 \, \textrm{Hz}$, Fig.s~\ref{fig:metaPlate}H and I, respectively), and above 
	($236 \, \textrm{Hz}$, Fig.~\ref{fig:metaPlate}J) the BG, with color corresponding to the acceleration magnitude, normalized by the maximum value in each field. 
	The fields for frequencies inside the BG 
	reveal significant attenuation in $\resDom$ --- identified as patches of light color --- compared to $\prisDom$, denoting strong activation of the resonators. In contrast, frequencies outside of the BG show nearly symmetric response between $\prisDom$ and $\resDom$, with high acceleration on both sides, except at nodal lines.
	
	Next, we 
	characterize the drop response regimes 
	that can be activated at different frequencies, with special attention to any dichotomy that may arise between $\prisDom$ and $\resDom$ due to the 
	resonators. We recognize that the plate features a spatially heterogeneous response, as shown in Fig. \ref{fig:pristPlate}, which may persist within each region, resulting in an unavoidable spatial variability that may partially obscure the dichotomy due to the LRM that we want to reveal. To filter out this variability, we subdivide $\prisDom$ and $\resDom$ into $9 \times 5$ subsections, as shown in Fig. \ref{fig:metaPlate}K. Each subdomain is tested and ranked for its ability to activate a drop and set it into motion from a pinned state. We deposit a 20 \unit{\micro \liter} drop (water + rhodamine B dye) in a subdomain and apply a harmonic excitation for 60 seconds, visually checking whether drop motion is triggered. 
	We repeat this exercise for each subdomain, and for a variety 
	of frequencies inside and outside the BG. 
	To classify the outcomes, we use the 
	binary labeling system: \emph{red} for a drop that remains pinned throughout the vibration cycle, and \emph{green} for a drop that depins and slides in that interval. Fig. \ref{fig:metaPlate}L illustrates a sample scenario 
	featuring a highly dichotomous outcome: drops deposited on the top row \emph{a} of $\prisDom$ (blue outline), stationary 
	at $t=0$, move under excitation --- sliding down the incline for $t>>0$ --- and are assigned a green label, while 
	drops deposited on the top row \emph{a} of $\resDom$ (orange outline), remain stationary and are assigned a red label.
	
	We aggregate the results in the binary grid plots of Fig. \ref{fig:metaPlate}M--T, where, at each frequency, we show 
	a $9 \times 5$ table of red/green pixels, 
	according to the inferred drop motion status 
	(on or off) for drops \emph{starting} at those subdomains. 
	We consider frequencies below the BG ($50 \, \textrm{Hz}$, Fig.s~\ref{fig:metaPlate}M,N), inside the BG ($100 \, \textrm{Hz}$ and $125 \, \textrm{Hz}$, Fig.s~\ref{fig:metaPlate}O,P and Q,R, respectively), and above the BG 
	($200 \, \textrm{Hz}$, Fig.s~\ref{fig:metaPlate}S,T). We accompany these results with $9 \times 5$ grids 
	containing the accelerations experienced by the individual subdomains, inferred from laser Doppler vibrometry measurements. 
	The results indicate with overwhelming evidence that, indeed, for excitations in the BG, 
	the attenuation in $\resDom$ prevents drops from sliding, while drops in $\prisDom$ almost universally move, with the exception of a few outlier locations. Outside the BG, in contrast, 
	$\prisDom$ and $\resDom$ show a more comparable response. Note that some 
	inertial effects due to the mass of the resonators, which elude the frequency sensitivity of the BG, 
	persist and continue to slightly attenuate the response in $\resDom$, even outside the predicted attenuation interval. Specifically, at $200 \, \textrm{Hz}$, the presence of nodal lines in the acceleration profile of both domains can be invoked to  explain the residual heterogeneity of response. 
	
	To provide a 
	rationale for the dichotomous behavior observed in the BG 
	with and without resonators, we complement our analysis with high-speed camera captures of a drop vibrating at $125 \, \textrm{Hz}$ in $\prisDom$ and $\resDom$, shown in Fig. \ref{fig:metaPlate}U. For each region, we extract the profile of the undeformed drop $\profile$ and compare it against that of the deformed drop, labeled $\profile_P$ in $\prisDom$ and $\profile_R$ in $\resDom$. Comparing the difference in deformation $\Delta$, we find that the drop in $\prisDom$, indeed, deforms more 
	than that in $\resDom$ --- i.e., $\Delta(\profile,\profile_P) > \Delta(\profile,\profile_R)$.

	\begin{figure}[t]
		\centering
		\includegraphics{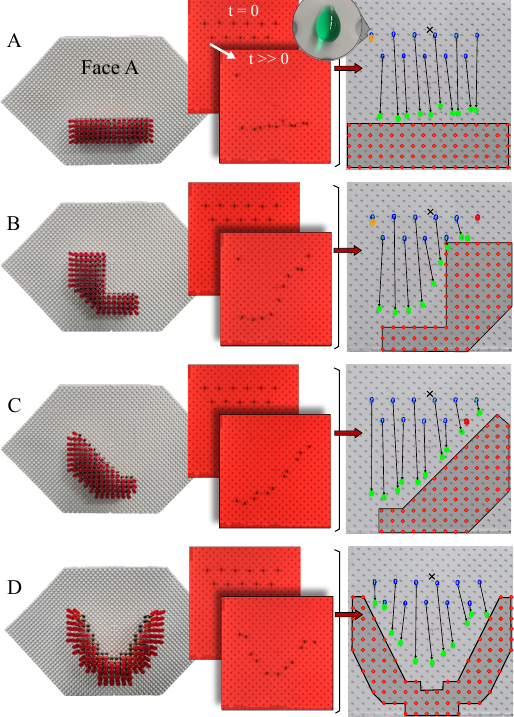}
		\caption{Drop motion and clustering programming using resonator patches with different patterns. The regions with resonators experience attenuated vibrations for frequencies in the BG, and act as road blocks for sliding drops, forcing them to aggregate at their edges. Examples shown include the following resonator patterns: (A) flat rectangle, (B) steps, (C) ramp, and (D) fork. For all cases, the drops cluster precisely along the patch edges. The experiments reveal remarkable ability to program the shapes of drop clusters according to a nearly endless variety of configurations, enabled by the assembly versatility of the LEGO\lego{} platform.}
		\label{fig:dropApplications}
	\end{figure}
	
	We can exploit these results to design a variety of LRM configurations that impart spatially-selective control patterns to a population of droplets deposited on a surface. Such spatial control can be used to achieve 
	drop \textit{clustering} along arbitrary \textit{aggregation edges}. To this end, we leverage the fact that a moving drop can be brought to a full stop when it encounters an array of resonators along its motion path due to attenuation. 
	We explore the variability of aggregation patterns that can arise when 
	drops interact with resonator arrays 
	of different shapes and sizes, and to what extend this interaction can be engineered. For this application, we take full advantage of the reconfigurability of the 
	LEGO\lego{} LRM platform, which allows for facile reorganization 
	of the resonators, enabling on-demand programming of the plate attenuation characteristics. 
	
	We organize resonators into 
	patches of different shapes. 
	Sample configurations are shown in Fig.~\ref{fig:dropApplications}: A) flat rectangle, B) steps, C) ramp, and D) fork. We deposit 20 \unit{\micro \liter} water drops in a scatter across the top of the PDMS substrate 
	and apply a sinusoidal excitation at $125 \, \textrm{Hz}$ (in the BG), which forces the drops to slide towards the resonators patch, 
	keeping it sustained until all drops have come to a stop. 
	We track the motion of each drop, 
	dyed green, by conducting the experiments under red light to increase their 
	contrast with the background for image processing. The drop profiles are extracted at the initial and final frame, and superimposed on a single image to highlight the trajectories (marked by arrows). At their initial positions the drop contours are highlighted in blue, while at their final positions they are in green if they stop at the predicted location (the patch edge), red if they do not move at all, and orange if they move but do not stop at the patch edge. 
	The results overwhelmingly show that drops slide down the incline up until the point where they encounter the resonators, which attenuate the vibration preventing motion beyond them. This results in drop aggregation landscapes that precisely conform to the shapes of the patch edges. Since the drops move on a pristine face, where we have no direct observation of the resonators, the aggregation pattern can be interpreted as providing a sort of \textit{hydrodynamic imaging} of the LRM configuration. 

	In conclusion, 
	we have shown unprecedented ability to achieve spatially-selective programming of drop motion, exploiting the interplay between flexible vibrating substrates and locally resonant metamaterials. 
	Ultimately, this capability allows us to program numerous landscapes of drop clustering patterns according to a nearly endless set of resonator configurations. 
	
	The authors acknowledge the support of the National Science Foundation (NSF grant CMMI-2211890).
	
	\bibliography{references.bib}
	
\end{document}


\title{Supplemental Materials: Droplet motion programming using metamaterials}
	\author{Mohammad Charara}
	\affiliation{Department of Civil, Environmental, and Geo- Engineering, University of Minnesota, Minneapolis, Minnesota 55455, USA}
	\author{Zak Kujala}
	\author{Sungyon Lee}
	\affiliation{Department of Mechanical Engineering, University of Minnesota, Minneapolis, Minnesota 55455, USA}
	\author{Stefano Gonella}
	\affiliation{Department of Civil, Environmental, and Geo- Engineering, University of Minnesota, Minneapolis, Minnesota 55455, USA}
	\email{sgonella@umn.edu}
	
	\maketitle

\section{Scanning Doppler Vibrometer Experiments}

3D Laser Doppler vibrometry (Polytec PSV-400-3D) experiments are performed to assess the out-of-plane response 
of points across the surface of the plate. Three lasers measure the velocity of select \emph{scan points} on the surface of the angled plate, while an electromechanical shaker (Br\"{u}el \& Kj\ae{}r Type 4810), attached to the plate through a superglued stinger, excites a signal generated and triggered by the vibrometry software through an amplifier (Br\"{u}el \& Kj\ae{}r Type 2718). The velocity data are decomposed into $\hat{x}$,$\hat{y}$, and $\hat{z}$ components by using Euler angles internally calculated by the vibrometry software during laser alignment and calibration. Prior to measurement, the surface of the baseplate is treated with a retroreflective spray (Reflect-All) reduce noise in the measurements and increase the signal reflected to the laser. We use retroreflective tape to increase the signal of desired scan points on the polydimathylsiloxane (PDMS) substrates deposited on the plate because the spray does not adhere to the PDMS surface. The plate was mounted at $45^\circ{}$ from the horizontal (to simulate the conditions for later drop motion testing) using angled clamps, and we enforced clamped boundary conditions on the left and right edges by sandwiching the baseplate between two pieces of wood on each side. The top and bottom edges were left open. In this work, we perform vibrometry measurements for two experimental exercises: 1) finding the bandgap and generating wavefields for the entire plate, and 2) generating max acceleration grids for the subdomains in $\prisDom$ and $\resDom$.

To plot the frequency response curve and generate the wavefield plots, 
we excite the baseplate with a 1--400 \unit{\hertz} broadband pseudorandom signal and collect surface velocity data from a dense grid of 1294 scan points across the entire surface of the baseplate, encompassing regions with and without deposited PDMS. During the scan, the results for each point are averaged over 15 readings to reduce the effect of spurious measurements. The internal vibrometry software performs a discrete Fourrier transform (DFT) on the collected data to generate frequency response curves of displacement, velocity, and acceleration for each scan point. We import the collected data into MATLAB. Focusing only on the out-of-plane acceleration response, we calculate the frequency response of $\prisDom$ and $\resDom$ by averaging the response of all points contained within the region, to obtain a single curve for each. The bandgap is roughly estimated as the gap between the two curves when plotted together. The wavefields are generated by spatially interpolating the amplitude of the frequency response at a chosen frequency over all scan points (cubic interpolation with MATLAB function \lstinline[style=Matlab-editor]{griddata}).

To generate the max acceleration grid for the individual $\prisDom$ and $\resDom$ subdomains, we excite the plate with a single-frequency sine signal and collect data from scan points in each subdomain -- here chosen to be three points in each subdomain on the PDMS substrates. The scan, averaged over 10 measurements for each point, collects the velocity of the surface of the plate at each scan point over 5-cycles of the response (i.e., temporal data). We import this data into MATLAB for processing. We use the low-pass filtering function \lstinline[style=Matlab-editor]{lowpass} to remove any components of the response signal $5 \times$ above the frequency of the actual excited signal to reduce any noise in the system that may have arisen from errors -- while the response of the system is nonlinear, higher harmonics are not expected to dominate the response so we do not expect this filters out useful data -- and perform three-point numerical differentiation in the temporal direction to calculate the acceleration of each scan point over the 5 cycles of data collected. We average the magnitude of the maximum and minimum acceleration across the five cycles to calculate the maximum acceleration at a scan point, and average this maximum over the three scan points in each subdomain. Data is plotted on a $9 \times 5$ grid for $\prisDom$ and $\resDom$, where the color of each box in the grid represents the acceleration amplitude at the corresponding subdomain.

All vibrometry experiments are performed with the plate at 45$^{\circ}$ to simulate experimental conditions experienced by the drops. Nevertheless, these results hold for a plate at an arbitrary angle. The experimental setup is shown in Fig. \ref{fig:pristPlate}A. 

\section{High-Speed Drop Motion Image Captures}

A baseplate, featuring a single symmetrically deposited PDMS patch on face B, is clamped at a 45$^\circ{}$ from the horizontal, and an electromechanical shaker is attached via a superglued stinger. A 20 \unit{\micro \liter} drop is deposited on the surface of the PDMS at a specific location using a 10-100 \unit{\micro \liter} precision pipette. A high-speed camera (Photron AX200) equipped with a macro lens is positioned on one side of the plate, as close as possible to the drop, such that it captures the side profile of the drop. A diffuse LED light panel is positioned opposite to the camera and pointing at the lens, acting as a background ligting throughout the experiment. A single-frequency signal is excited into the plate by the shaker through an amplifier, while the high-speed camera records the vibrating drop at 2000 frames per second. The vibrating drop is recorded for a few seconds or [in the case it depins] until it has moved out of the frame, whichever comes first, and the first about 1000 frames of the vibration, including the onset of deformation, are extracted and saved for further processing. This process is repeated for all plate locations of interest, with each exercise performed three times.

For each video, we find a time frame when the cyclic drop deformation has reached steady-state and extract a full drop deformation cycle by isolating the associated frames. These frames are imported into MATLAB for further processing, where the contrast is increased to more easily identify the drop profile. The drop's profile is computationally extracted at all frames during the deformation cycle, and for a single frame of the sessile drop prior to vibration, such that each profile is defined by a curve. With the increased contrast, the drop appears dark against the light backround and we define the boundary pixel between the dark and light areas as the drop profile. We quantify the deformation of the drop by calculating the area difference between the deformed and and undeformed drop profiles, and normalize it by the area of the undeformed drop. For each drop, we consider the maximum deviation from the undeformed case for the linear fit in Fig. \ref{fig:pristPlate}S.

\section{Drop Motion Experiments: Drop Motion Capability in Subdomains}

A baseplate, featuring a pattern of resonators on face A and two symmetrically deposited PDMS patches on face B, is clamped at a 45$^\circ$ from the horizontal and an electromechanical shaker is attached to its center via a superglued stinger. Domains within each PDMS patches, identified as $\prisDom$ and $\resDom$, are divided into $9 \times 5$ subdomains each. Drops 20 \unit{\micro \liter} in volume, consisting of a water and fluorescent dye mixture, are deposited on each subdomain using a 10-100 \unit{\micro \liter} precision pipette. A single-frequency sine signal is excited in the plate via a signal amplifier and sustained for 60 seconds. The experiment is recorded via a camera, and we track drops throughout. We repeat for all subdomains. For convenience and efficiency, experiments are conducted placing drops across an entire row in each subdomain (5 columns each), essentially extracting the results for 10 drops (5 for $\prisDom$ and $\resDom$) at once. 
The experiments are conducted in darkness under green light as the fluorescent dye glows in these conditions. This makes image processing easier as it simplifies the extraction of the drop profiles computationally. We record results on a $9 \times 5$ table that characterizes the behavior of a drop initially deposited at the corresponding location in each domain in a binary fashion: motion/no motion. Comparing a frame before the start of excitation to a frame after the end of excitation, a drop that moves from its initial location (i.e., the drop depins resulting in motion of its entire profile) results in labeling the corresponding subdomain green, while a drop that remains pinned labels the subdomain red. In the case of drops that exit the frame before the end of the experiment or drops that coalesce, we track the motion of the drop throughout the video to make our assessment. Although drops that start in certain subdomains move very far from their starting point by the end of their experiment, while others move very little, we do not concern ourselves with how far or how fast they move, only if they translate.


\bibliography{references.bib}